# Compact Modeling of *I-V* Characteristics, Temperature Dependency, Variations, and Noise of Integrated, Reproducible Metal-Oxide Memristors

H. Nili[1], A. Vincent[1], M. Prezioso[1], M. R. Mahmoodi[1], I. Kataeva[2] and D. Strukov[1*]

[1] UC Santa Barbara, Santa Barbara, CA 93106-9560, U.S.A. email: *strukov@ece.ucsb.edu
[2] Research Laboratories, DENSO CORP., 500-1 Minamiyama, Komenoki-cho, Nisshin, Japan 470-0111

*Abstract*---We present a comprehensive phenomenological model for the crossbar integrated metal-oxide continuous-state memristors. The model consists of static and dynamic equations, which are obtained by fitting a large amount of experimental data, collected on several hundred devices. The static equation describes the device current, at non-disturbing voltages, as a sum of three components, representing the average behavior and its device-to-device and temporal variations. All three components are expressed as functions of the applied voltage, ambient temperature, and the current memory state. The dynamic equation models the change in the memory state as a function of the applied voltage stress and the current memory state, and is also expressed as a sum of the average and the device-to-device variation terms. Both equations are explicit, computationally inexpensive, and suitable for SPICE modeling. At the same time, the model has good predictive power, which is supported by the validation results. The presented model is useful for realistic simulations of various mixed-signal computing circuits, such as image compression and image classification applications, discussed in this paper.

## I. Introduction

Though there has been significant progress in understanding the physics of operation for the metal-oxide ReRAM devices (also called memristors), the development of accurate and comprehensive compact models has proven to be still very challenging. For the devices based on strongly-correlated materials, this is in part due to very rich transport and state-changing physics, because multiple mechanisms, e.g., switching due to ferroelectric domains, metal insulator transition, and drift of the defects, can be involved in the device operation [1].

Indeed, most reported work (e.g. [2-4]) was based on physics models. This is very useful for verifying physical mechanisms, and as a guidance towards engineering of better devices. However, such models are generally inadequate for accurate and fast simulations, in part because of their focus on just certain aspects of the device operation, but also due to their implicit form, such as systems of coupled differential equations. The reported compact and SPICE models [5-8] are not sufficiently detailed (e.g., lacking device variations, temperature dependences) and/or accurate, largely because they are derived based on simplified assumptions about resistive switching and electron transport mechanisms.

The need for accurate and comprehensive compact model of memristors is very acute now, given the recent advances in this technology and the increased focus on investigating memristors' potential for various applications. The main contribution of this paper is the development of such a model using a phenomenological approach.

## II. Integrated Metal-Oxide Memristors

Our most detailed model was developed for $Pt/Al_2O_3/TiO_{2-x}/Ti/Pt$ devices, integrated in the 20×20 crossbar arrays, with 200-nm lines separated by 400-nm gaps (Fig. 1). Details of this fabrication technique, as well as extensive endurance and retention data, have been reported in [9]. The studied crossbar devices have uniform *I-V* characteristics with a narrow spread of set and reset voltages (Fig. 1c) - a critical requirement for the model development. The same approach to modelling has been also verified for two other types of crossbar-integrated memristors.

## III. General Modeling Approach

We model the device behavior with two equations: $I = S(G_0, V, T)$ and $\Delta G_0 = D(G_0, V_p, t_p)$. The first, "static" equation describes the device current ($I$) as a function of the applied voltage ($V$), ambient temperature ($T$), and its memory state (represented by its low-voltage $G_0 \equiv I(0.1V)/0.1V$), at relatively small voltage biases (in our case, below 0.4 V) that do not modify the state, within the studied range of temperatures. The second, "dynamic" equation describes the change of the device's memory state as a function of its initial state after the application of a voltage pulse with some amplitude $V_p$ and duration $t_p$. Such method of separating the static and dynamic behavior works well for most practical devices with strongly-nonlinear switching kinetics.

These two equations are expressed as $S = S_m(G_0, V, T) + S_{d2d}(G_0, V, T) + S_N(V)$ and $D = D_m(G_0, V_p, t_p) + D_{d2d}(G_0, V_p, t_p)$, where $S_m$ and $D_m$ represent, respectively, the device's expected noise-free dc *I-V* curves and the expected conductance change $\Delta G$, while $S_{d2d}$ and $D_{d2d}$ are their normally-distributed stochastic d2d variations, and $S_N(V)$ is the temporal variations due to device's intrinsic noise.

## IV. Modelling Results

To model the static behavior, we recorded dc *I-V* curves in the range -0.4 V to +0.4 V from 324 devices at 6 different temperatures: RT, 40, 55, 70, 85 and 100 °C. All the curves have been then fitted with a cubic polynomial of the form $A_1(G_0, T)V + A_3(G_0, T)V^3$, mainly due to its simplicity. Subsequently, the $A_1$ and $A_3$ values are separately fitted to find the parameters $\mu_{A1}$ and $\mu_{A3}$, (Fig. 2a, b) and $\sigma_{A1}$ and $\sigma_{A3}$ (Fig. 2c, d) that describe, respectively, the averages of $A_1$ and $A_3$, and their effective standard deviations as functions of temperature and conductance. To make the model more representative, 22 outlier devices for which the parameter $A_1$ deviated by more than 50% from a quadratic surface fitting were excluded from the fitting set (Fig. 1d). Following this approach, the first two terms of the static equation can be written as $S_m = \mu_{A1} V + \mu_{A3} V^3$, and $S_{d2d} = N(0,1) (\sigma_{A1} V + \sigma_{A3} V^3)$. A simple conductance-state-dependent noise model was fitted with experimental data (Fig. 3), which are comparable with the previous work [10].

The same 324 devices were used for deriving the dynamic model. In this case, all devices were first randomly initialized to memory states within the 3.5 μS to 300 μS range. Each device was then subjected to 6,000 voltage pulses with random polarity and amplitude (from the ranges [-0.8 V, -1.5 V] / [0.8 V to 1.15 V] for set / reset) and duration (from 100 ns to 100 ms), resulting in a total dataset size of ~2 million points that is covering uniformly the whole range of memory states. To avoid irreversible damage, the devices were not stressed beyond the studied range of conductances. Functions $D_m$ and $D_{d2d}$ were found separately for 8 ranges of memory states. For each range, $D_m$ was found by fitting the data of all devices in each range to the empirically chosen exponential functions (Fig. 4). To obtain $D_{d2d}$, the coefficients of variation (CV) for conductance change were first calculated from the experimental data (Fig. 5), and then fitted with a polynomial function of $V_p$ and log $t_p$.

The functional forms and all fitting parameters for the baseline device technology are summarized in Fig. 6.

The accuracy of the static model was first validated using the R-squared measure, calculated for all the experimental I-Vs and the ones defined by the $S_m$ model (Fig. 7). Fig. 8 shows that the model can predict the average static behavior and d2d variations at different temperatures very well. The accuracy of the dynamic model was partially verified by contrasting the statistics of the experimentally observed conductance changes under different applied voltage pulses, and the ones predicted by the models (Fig. 9a,b). Finally, Fig. 9c,d shows the simulated evolution of conductance (at 0.1 V) for 5 devices, demonstrating the role of pulse stress conditions and device-to-device variations in the dynamic behavior.

## V. DISCUSSION AND SUMMARY

Although the model has been developed for a particular (though representative) fabrication technology, this approach to memristor modelling is quite general and can be applied to other types of such devices. For example, Fig. 10 shows preliminary results for two other device stacks (fabricated using a somewhat different technology), for which similar polynomial functions, but with different fitting parameters, were successfully used to model the static behavior.

The developed model is useful for guiding the design automation for large analog and mixed-signal computing circuits based on integrated analog-state memristors. To demonstrate that, we have simulated image compression and reconstruction circuit employing the discrete cosine transform (DCT) algorithm, and compared modeling results with experimental ones. The DCT algorithm, which is similar to the one considered in [11], involves the vector-by-matrix multiplication (VMM), which can be implemented with 8×8 analog ReRAM crossbars. Fig. 11e shows that the simulated results for pixel intensities of the reconstructed image follow closely the experimentally measured data. Additionally, the model was applied to simulate the impact temperature variations and device imperfections on the functional performance of ex-situ trained medium-scale image classifiers (Fig. 12).

It is worth noting that the choice of the fitting functions may be guided by plausible physical mechanisms as well as computational intensity. The polynomial function used in the static equation fitting had been indeed motivated by a plausible conduction model, with the linear part representing the Ohmic conductance of vacancy-doped filaments, while the nonlinear one approximating the bulk trap-limited conduction of the insulating parts of the metal-oxide layer and/or the charge-carrier injection at the interfaces.

In summary, our paper presents a comprehensive compact model for metal-oxide memristors, which is developed via empirical fitting of the measured data. The model includes static and dynamic characteristics, their device-to-device variations, temperature effects, and noise. The functionality of the model has been demonstrated using two case studies of practical memristive networks.

ACKNOWLEDGMENTS

This work was supported by NSF grant CCF 1740352, SRC nCORE NC-2766-A, and DENSO CORP., Japan. The authors are grateful to B. Chakrabarti, F. Merrikh Bayat, and especially K. K. Likharev for useful discussions and technical support.

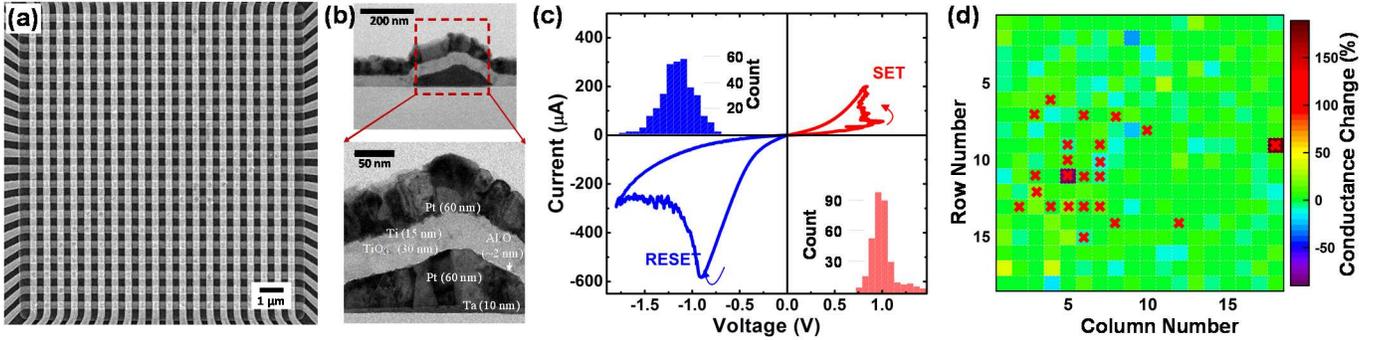

**Fig. 1.** Reproducible integrated metal-oxide memristors: (a) Top view SEM image of a 20 × 20 bilayer metal-oxide crossbar. (b) Cross-sectional TEM view of a single device in the crossbar array, with different metal and oxide layers identified. (c) Representative *I-V* characteristics and set-reset threshold statistics. (d) Conductance state retention (at 0.1 V) at the process of taking the *I-V* data for static modelling, at small non-disturbing biases (< 0.4 V), and temperatures up to 100 °C. Crosses denote the devices removed from fitting set, due to abnormal behavior.

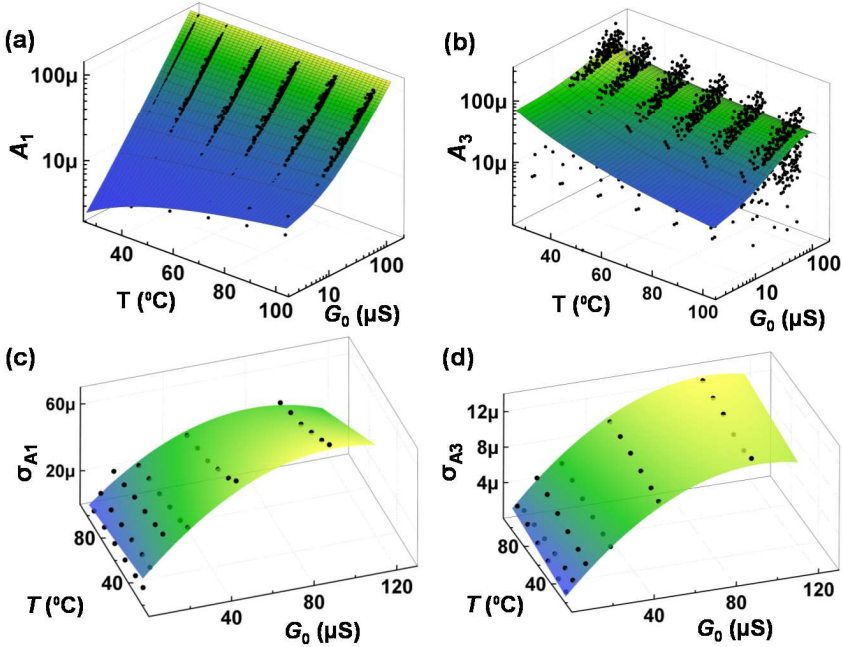

**Fig. 2.** Static model: (a, b) the data (dots) and fitting surfaces for the model parameters $A_1(G_0,T)$ and $A_3(G_0,T)$ and (c, d) their corresponding d2d variations. Specifically, in panels (c, d) the dots show standard deviations calculated for the $A_1$ and $A_3$ values in each bin, while the surfaces show their fitted functions.

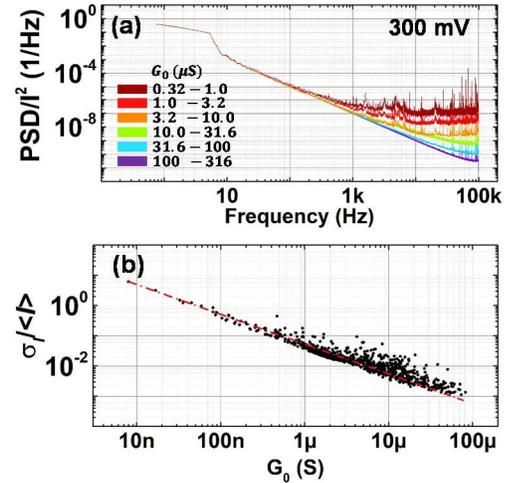

**Fig. 3.** Noise model: (a) The normalized power spectral density ($S_I/I^2$) measured at the reading voltage 0.3 V. The spectra have been collected for all 324 devices, binned into 6 conductance ranges and averaged. The corner frequency is lower for states with lower conductance, because of the noise floor of the measurement setup. (b) The coefficient of variation of the current time series data collected from all the devices at 0.1 V, 0.2 V and 0.3 V, shown as a function of $G_0$. The data were collected for 1 s at a sampling rate of 200 kHz.

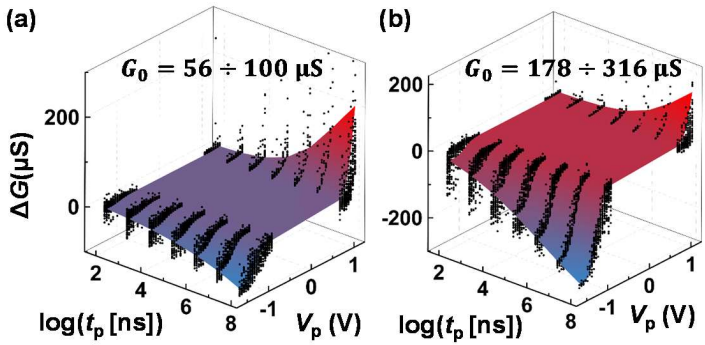

**Fig. 4.** Dynamic model: The measured absolute change in conductance at 0.1 V as a function of the amplitude ($V_p$) and duration ($t_p$) of the applied voltage pulse for two state ranges (out of 8 total used). On each panel, dots show the experimental data for the devices with the initial memory states in that particular range. The surfaces show the corresponding fitted functions.

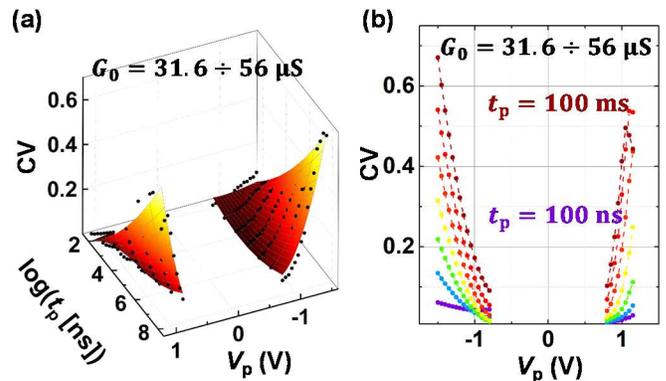

**Fig. 5.** Dynamic model for device-to-device variations: (a) 3D and (b) 2D graphs of the coefficients of variation and their polynomial fits for a representative conductance range.

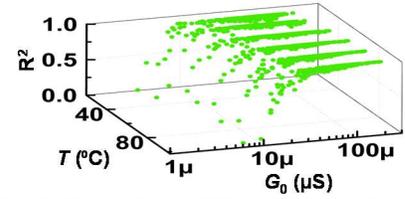

Fig. 7. The goodness-of-fitting ($R^2$ measure) obtained with $S_m$ function applied to all experimental I-V curves.

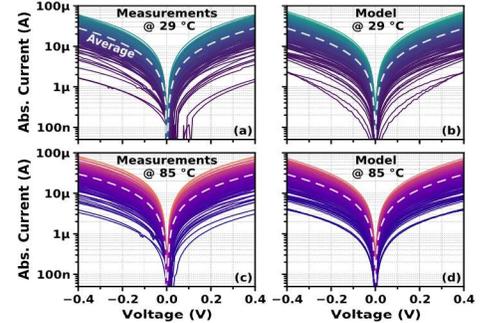

Fig. 8. Comparison between the experimental I-V curves measured from 200 random devices and the simulated I-V curves based on $I = S(G_0,V,T)$, using a similar set of initial conductances at 0.1 V at 29° and 85° C.

Fig. 6. Summary of the model and its parameters for Pt/Al$_2$O$_3$/TiO$_x$/Ti/Pt crossbar devices.

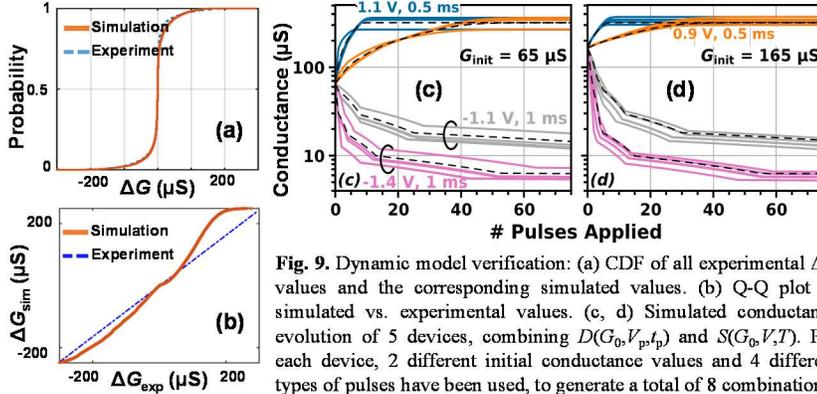

Fig. 9. Dynamic model verification: (a) CDF of all experimental $\Delta G$ values and the corresponding simulated values. (b) Q-Q plot of simulated vs. experimental values. (c, d) Simulated conductance evolution of 5 devices, combining $D(G_0,V_p,t_p)$ and $S(G_0,V,T)$. For each device, 2 different initial conductance values and 4 different types of pulses have been used, to generate a total of 8 combinations.

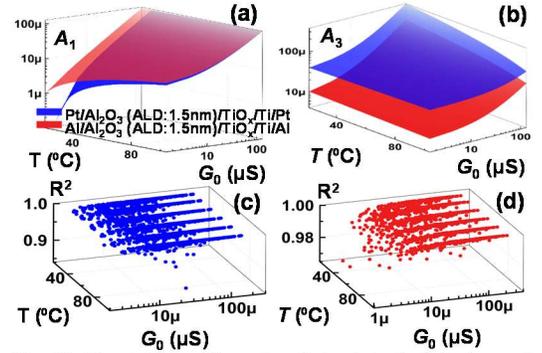

Fig. 10. The static model results, obtained similarly to those of Figs. 2 and 7 for two other types of studied devices (with different deposition approach and electrode material).

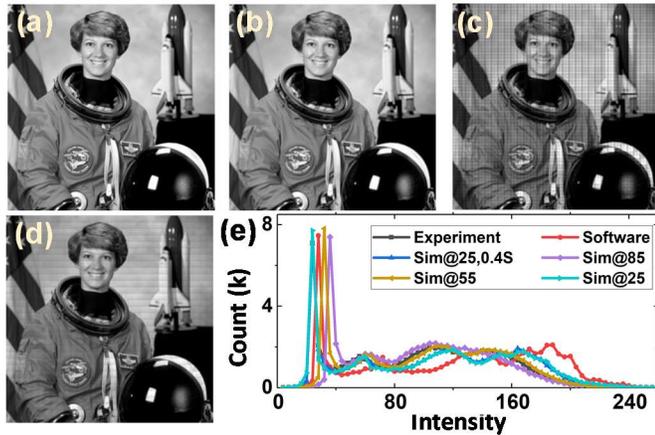

Fig. 11. Compressing and decompressing an image using DCT algorithm: (a) an original and (b-d) reconstructed images, obtained by (b) device-oblivious full precision simulations, (c) experimentally measured data, and (d) using the proposed model. (e) The histograms of the pixel intensities for reconstructed images. The standard deviation for the pixel intensity differences between (c) and (d) images, normalized to the maximum brightness, is less than 5%.

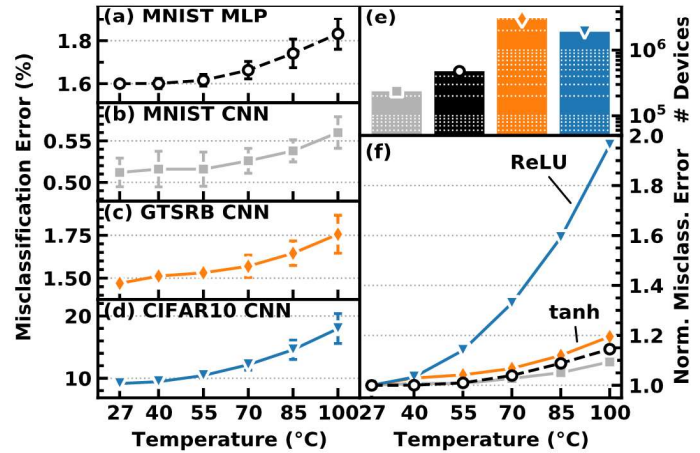

Fig. 12. Application of the model for simulation of ex-situ trained mixed-signal image classifiers for (a,b) MNIST, (c) GTSRB, and (d) CIFAR10 benchmarks. The neural network parameter and training procedure is similar to those used in [12, 13]. (e) The number of memristors utilized in each simulated network. (f) The misclassification error for considered cases, normalized by its value at 27° C.